\documentclass[12pt]{article}

\usepackage{amsmath}
\usepackage{amssymb}
\usepackage{latexsym}
\usepackage{graphics}
\usepackage{psfrag,fancyhdr,epsfig}

\begin{document}

\begin{titlepage}

\hfill hep-th/0611181\\

\vspace*{1.5cm}
\begin{center}
{\bf \Large Brane Cosmology with a Non-Minimally\\
 Coupled Bulk-Scalar Field}

\bigskip \bigskip \medskip

{\bf  C. Bogdanos, A. Dimitriadis and K. Tamvakis}

\bigskip

{ Physics Department, University of Ioannina\\
Ioannina GR451 10, Greece}

\bigskip \medskip
{\bf Abstract}
\end{center}
We consider the cosmological evolution of a brane in the
 presence of a bulk scalar field coupled to the Ricci scalar
  through a term $f(\phi)R$. We derive the generalized Friedmann equation
   on the brane in the presence of arbitrary brane and bulk-matter, as well as the scalar field equation, allowing
   for a general scalar potential $V(\phi)$. We focus on a quadratic form of the above 
   non-minimal coupling $-\xi\phi^2R$
    and obtain a class of 
   late-time solutions for the scale factor and the scalar field on the brane that exhibit accelerated
    expansion for a range of the non-minimal coupling parameter $\xi$.

\end{titlepage}

\section{Introduction}

In the last few years, an increasing number of cosmological
data~\cite{R},\cite{WMAP} show that the universe undergoes accelerated
expansion attributed to an energy component referred to as {\textit{``dark
energy"}}. Dark energy is a significant percentage of the total energy of the
universe. Although a cosmological constant is admittedly the simplest model  of
dark energy, the huge fine-tuning required for its magnitude makes theorists
unhappy~\cite{COSMO},\cite{Nesseris}. Other attempted explanations of the origin of dark energy
are phantom fields~\cite{PHANT} and quintessence~\cite{QUINT}, as well as
modifications of gravitational theory itself~\cite{DGP}. Independently of the dark energy puzzle,
 theories of extra spatial dimensions,
in which the observed universe is realized as a {\textit{brane}} embedded in a
higher dimensional spacetime ({\textit{bulk}}), have received a lot of attention
in the last few years. In this framework ordinary matter is trapped on the brane
but gravitation propagates throughout the entire
spacetime~\cite{A},\cite{AHDD},\cite{RS}. The cosmological evolution on the
brane is described by an effective Friedmann equation~\cite{SMS},\cite{BDL} that
incorporates non-trivially the effects of the bulk.  Brane models provide us
with new possibilities for the understanding of open cosmological issues like
the observed accelerated expansion. The existence of a higher dimensional
embedding space allows for the existence of bulk matter, which can certainly
influence the cosmological evolution on the brane and be a major contributor to
the dark energy. A particular form of bulk matter is a scalar field\cite{Langlois}. The
presence of a bulk scalar field opens up the possibility of a direct coupling of
to the curvature scalar. A specific form of this coupling corresponds
to the gravitational term appearing in the so-called tensor-scalar theory of
gravity~\cite{BD}. A bulk scalar field non-minimally coupled via a coupling of
the form $f(\phi)R$ has also been considered in the Randall-Sundrum framework
and a class of analytic as well as numerical solutions have been
discussed~\cite{KT},\cite{FP},\cite{Barbosa}. 

In the present article we study the cosmological evolution on a brane in the
presence of a bulk scalar field, non-minimally coupled to the Ricci scalar \footnote{A similar case with a 4-D scalar non-minimally coupled to induced Ricci curvature was treated in \cite{WANDS}.}. We derive the
cosmological evolution equation on the Brane as well as the
corresponding scalar field evolution equation. In the latter we allow for a general scalar potential function.
In addition to the scalar potential, these equations contain the unspecified function $\hat{\phi}''(t)$, 
standing for the
non-distributional part of the second derivative of the scalar field
with respect to the fifth spatial coordinate. We focus on a quadratic choice of the coupling to the Ricci
scalar, namely $f(\phi)=2M^3(1-\xi\phi^2/2)$. Assuming a simple quadratic form for the scalar potential on the brane
and employing various
ans\"atze for $\hat{\phi}''(t)$, we derive a class of approximate late-time
solutions. These solutions exhibit accelerated expansion for a range of
the non-minimal coupling parameter $\xi$. 
The plan of the paper is the following. In Section 2 we introduce the general
framework of the model and derive Einstein's
equations for arbitrary bulk and brane-matter. In Section 3 we calculate the
evolution equations on the brane for a general coupling function. We focus on the
 specific case of a quadratic non-minimal coupling $-\xi\phi^2\,R$ and
write down the equations in a {\textit{``dark energy variable''}} formulation. In Section 4 we discuss the 
late time behaviour in the absence of matter. Suitable choices for $\hat{\phi}''$ and the scalar potential
 give accelerated expansion driven by a cosmological constant and a scalar field 
exponentially falling in time. Finally, in Section 5,
we present two approximate late time solutions for two different
ans\"atze of $\hat{\phi}''$ exhibiting accelerated expansion for a limited range of the coupling parameter $\xi$.

\section{General Framework}

Consider the Action
\begin{eqnarray}
{\cal{S}}=&\int &\,d^5x\,\sqrt{-G}\,\left\{\,f(\phi){\cal{R}}\,- 
\Lambda\,-\frac{1}{2}\left(\nabla\phi\right)^2\,-V(\phi)\,-{\cal{L}} 
_B^{(m)}\right\}\,\nonumber\\
\,\,+\,&\int &\,d^4x\,\sqrt{-g}\,\left\{\,-\sigma\,+\,{\cal{L}}_b^ 
{(m)}\,\right\}\,,{\label{Action}}
\end{eqnarray}
describing $5D$ gravitation in the presence of a scalar field $\phi$  
that couples non-minimally to the
$5D$ Ricci scalar $R$ through a general coupling function $f(\phi)$.  
Standard Model matter, described by ${\cal{L}}_b^{(m)}$,
is confined on a {\textit{Brane}} located at $x_5\equiv y=0$.  
Additional bulk matter, distinct from the
bulk scalar field $\phi$, is described by ${\cal{L}}_B^{(m)}$. The $5D 
$ metric $G_{MN}$ has signature
$(-,\,+,\,+,\,+,\,+)$. With $g_{\mu\nu}$ we denote the $4D$ metric on  
the Brane. Finally, $\sigma$ is the (positive)
Brane-tension and $\Lambda$ is the $5D$
cosmological constant.
Varying the Action with respect to the metric we obtain Einstein's  
equations
\begin{equation}
f(\phi)\left(\,{\cal{R}}_{MN}\,-\frac{1}{2}G_{MN}{\cal{R}}\,\right)\,- 
\nabla_M\nabla_Nf(\phi)\,+\,G_{MN}\nabla^2f(\phi)\,=\,\frac{1}{2}T_ 
{MN}\,,{\label{EOM}}
\end{equation}
where $T_{MN}$ is the total energy-momentum tensor
\begin{equation}
T_{MN}\,=\,T_{MN}^{(\phi)}\,+\,T_{MN}^{(B)}\,+\,T_{MN}^{(b)}\,-G_{MN} 
\Lambda\,-G_{\mu\nu}\delta_M^{\mu}\delta_N^{\nu}\sigma\,\delta(y)\,.
\end{equation}
$T_{MN}^{(\phi)}$ stands for the scalar field part of $T_{MN}$, namely,
\begin{equation}
T_{MN}^{(\phi)}\,=\,\nabla_M\phi\nabla_N\phi\,-G_{MN}\left(\,\frac{1} 
{2}\left(\nabla\phi\right)^2\,+\,V(\phi)\,\right)\,,
\end{equation}
while the scalar field equation of motion is
\begin{equation}
\nabla^2\phi\,-\frac{dV}{d\phi}\,+\,{\cal{R}}\frac{df}{d\phi}\,-\frac 
{d\sigma}{d\phi}\delta(y)\,=\,0\,.{\label{scalareom}}
\end{equation}

Let us now introduce the metric ansatz\cite{BDL}
\begin{equation}
ds^2\,=\,-n^2(y,t)dt^2\,+\,a^2(y,t)\gamma_{ij}dx^idx^j\,+\,b^2(y,t)dy^2
\end{equation}
or
\begin{equation}
G_{MN}\,=\,\left(\begin{array}{ccc}
-n^2(y,t)\,&\,0\,&\,0\\
0\,&\,a^2(y,t)\gamma_{ij}\,&\,0\\
0\,&\,0\,&\,b^2(y,t)
\end{array}\right)\,.{\label{metric}}
\end{equation}
This metric corresponds to a Friedmann-Robertson-Walker geometry on  
the brane with a maximally symmetric $3$-geometry
$\gamma_{ij}$. We assume ${\cal{Z}}_2$ symmetry and $y$ takes values  
in the $[-\infty,\,+\infty]$ line. The functions
$n(y,t),\,a(y,t)$ and $b(y,t)$ are continuous with respect to $y$ but  
may have discontinuous first derivatives at the
location of the brane.

For the metric ansatz ({\ref{metric}}), the scalar field energy-momentum tensor is
\begin{equation}
T_{MN}^{(\phi)}\,=\,\left(\begin{array}{ccc}
\frac{1}{2}\dot{\phi}^2+\frac{n^2}{2b^2}{\phi'}^2+n^2V\,&\,0\,&\,\dot 
{\phi}\phi'\\
0\,&\,-a^2\gamma_{ij}\left[-\frac{1}{2n^2}\dot{\phi}^2+\frac{1}{2b^2} 
{\phi'}^2+V\right]\,&\,0\\
\dot{\phi}\phi'\,&\,0\,&\,\frac{1}{2}{\phi'}^2+\frac{b^2}{2n^2}\dot 
{\phi}^2-b^2V
\end{array}\right)\,,
\end{equation}
where, we assume that the scalar field depends only on the fifth  
coordinate and denote by $\phi'$ its derivative with
respect to it and by $\dot{\phi}$ its derivative with respect to  
time. The bulk and brane parts of the energy-momentum
tensor can be parametrized as
\begin{equation}
T_{MN}^{(B)}\,=\,\left(\begin{array}{ccc}
\rho_Bn^2\,&\,0\,&\,-n^2P_5\\
0\,&\,P_Ba^2\gamma_{ij}\,&\,0\\
-n^2P_5\,&\,0\,&\,\overline{P}_Bb^2
\end{array}\right)\,\,,\,\,\,\,\,\,\,\,\,T_{MN}^{(b)}\,=\,\frac{\delta 
(y)}{b}\left(\begin{array}{ccc}
\rho n^2\,&\,0\,&\,0\\
0\,&\,pa^2\gamma_{ij}\,&\,0\\
0\,&\,0\,&\,0
\end{array}\right)\,,
\end{equation}
where the {\textit{bulk energy}} and {\textit{momentum densities}} $\rho_B,\,P_B, 
\,\overline{P}_B$, as well as the {\textit{energy exchange function}}
$P_5$, are functions of time and of $y$, while
the
{\textit{brane energy}} and {\textit{momentum densities}} $\rho,\,p$ are  
functions of time.

Substituting the above ansatz and choosing {\textit{Gauss normal coordinates}}  
($b(y,t)=1$), we obtain the
equations of motion in the following form

$$3\left\{\,\left(\frac{\dot{a}}
{a}\right)^2 - n^2 \left( \,\frac{a''}{a}\, + \,\left(\frac{a'}{a} 
\right)^2\, \right) \,+ \,k\frac{n^2 }{a^2 } \,\right\}\,f\,\,-n^2 
\left\{\,f''\,+\,3\frac{a'}{a}f'\,\right\}\,+
\,3\frac{\dot{a}}{a}\dot{f}$$
\begin{equation}\,=\,\frac{1}{4}\dot{\phi}^2+\frac{n^2}{4}{\phi'}^2+ 
\frac{n^2}{2}V(\phi)\,+
\,\frac{n^2}{2}\rho_B\,+\,\frac{n^2}{2}\delta(y)\rho\,
+\,\frac{n^2}{2}\delta(y)\sigma(\phi)\,+\,\frac{n^2}{2}\Lambda\,, 
{\label{00}}
\end{equation}
$$\,$$
\begin{equation}3\left( {\frac{{n'}}
{n}\frac{{\dot a}}
{a} - \frac{{\dot a'}}
{a}} \right)\,f\,
-\dot{f}'\,+\,\frac{n'}{n}\dot{f}\,=\,\frac{1}{2}\dot{\phi}\phi'\,- 
\frac{n^2}{2}P_5\,,{\label{05}}\end{equation}
$$\,$$
$$3\left\{ {\frac{{a'}}
{a}\left( {\frac{{a'}}
{a} + \frac{{n'}}
{n}} \right) - \frac{1}
{{n^2 }}\left( {\frac{{\dot a}}
{a}\left( {\frac{{\dot a}}
{a} - \frac{{\dot n}}
{n}} \right) + \frac{{\ddot a}}
{a}} \right) - \frac{k}
{{a^2 }}} \right\}\,f\,-\frac{1}{n^2}\left\{\,\ddot{f}\,+\,\left(3 
\frac{\dot{a}}{a}\,-\frac{\dot{n}}{n}\right)\dot{f}\,\right\}\,$$
\begin{equation}
+
\,\left(3\frac{a'}{a}+\frac{n'}{n}\right)f'\,=\,\frac{1}{4}{\phi'}^2+ 
\frac{1}{4n^2}\dot{\phi}^2-\frac{1}{2}V(\phi)\,-\frac{1}{2}\Lambda\,
+\,\frac{1}{2}\overline{P}_B\,,{\label{55}}\end{equation}
$$\,$$
$$a^2 \gamma _{ij} \left\{ {\frac{{a'}}
{a}\left( {\frac{{a'}}
{a} + 2\frac{{n'}}
{n}} \right) + 2\frac{{a''}}
{a} + \frac{{n''}}
{n}} \right\}\,f\, + \,\frac{{a^2 }}
{{n^2 }}\gamma _{ij} \left\{ \frac{\dot{a}}
{a}\left( - \frac{\dot {a}}
{a} + 2\frac{\dot {n}}
{n} \right) - 2\frac{\ddot {a}}
{a}  \right\}\,f\,$$
$$- kf\gamma _{ij}\,+\,\gamma_{ij}\left\{\,-\frac{a^2}{n^2}\left[\, 
\ddot{f}\,+
\,\left(2\frac{\dot{a}}{a}\,-\frac{\dot{n}}{n}\right)\dot{f}\,\right]
\,+\,a^2\left[\,f''\,+\,\left(2\frac{a'}{a}
+\frac{n'}{n}\right)f'\,\right]\,\,\right\}$$
\begin{equation}\,=\,-\frac{a^2}{2}\gamma_{ij}\left[-\frac{1}{2n^2} 
\dot{\phi}^2+{\phi'}^2+V(\phi)\right]\,+\,\frac{a^2}{2}\gamma_{ij} 
(P_B-\Lambda)\,
+\,\frac{a^2}{2}\gamma_{ij}\delta(y)\left(p-\sigma(\phi)\right)\,, 
{\label{ij}}\end{equation}
$$\,$$
\begin{equation}
\ddot{\phi}\,+\,\left(3\frac{\dot{a}}{a}-\frac{\dot{n}}{n}\right)\dot 
{\phi}\,
-n^2\left\{\,\phi''\,+\,\left(\frac{n'}{n}+3\frac{a'}{a}\right)\phi'\, 
\right\}\,
+n^2V'\,-n^2{\cal{R}}f'+n^2\sigma'\delta(y)=0\,.{\label{scalareq}}\end 
{equation}
The Ricci scalar ${\cal{R}}$ appearing in the scalar equation is
$${\cal{R}}=3\frac{k}{a^2}\,+\,\frac{1}{n^2}\,\left\{\,6\frac{\ddot 
{a}}{a}\,+\,6\left(\frac{\dot{a}}{a}\right)^2\,-6\frac{\dot{a}}{a} 
\frac{\dot{n}}{n}\,\right\}\,
-6\frac{a''}{a}\,-2\frac{n''}{n}\,-6\left(\frac{a'}{a}\right)^2-6\frac 
{a'}{a}\frac{n'}{n}\,$$
We have denoted with dots the derivatives with respect to time and  
with primes the derivatives with respect to the
fifth coordinate. Note though that $V'$ denotes a derivative with  
respect to the scalar field $\phi$. We have also allowed for a
$\phi$-dependence of the brane-tension $\sigma$ and denoted with $ 
\sigma'$ its derivative with respect to the scalar
field.

Assuming ${\cal{Z}}_2$ symmetry and denoting $a'(t)\equiv a'(+0,t), 
\,n'(t)\equiv n'(+0,t)$ and $\phi'(t)\equiv
\phi'(+0,t)$, we may proceed to extract from the equations of motion  
the {\textit{Junction Conditions}} at $y=0$. Thus,
from ({\ref{00}}), ({\ref{ij}}) and ({\ref{scalareq}}) we obtain
$$-6f\frac{a'}{a}-2f'\phi'=\frac{1}{2}(\rho+\sigma)$$
$$4f\frac{a'}{a}+2n'f+2f'\phi'=\frac{1}{2}(p-\sigma)$$
$$-2\phi'+4f'\left(n'+3\frac{a'}{a}\right)=-\sigma'\,$$
Note that, from now on, $f'$ {\textit{stands for a derivative with  
respect to}} $\phi$. We have also chosen $n(0,t)=1$.
These equations can be put in the form
\begin{equation}
\frac{a'}{a}=-\frac{1}{2}\frac{(\sigma+2f'\sigma')}{u}\,+\,\frac{1}{8} 
\frac{(3p-\rho)}{u}-\frac{1}{16}\frac{(\rho+p)}{f}\,,
{\label{Delta a}}
\end{equation}
\begin{equation}
n'=-\frac{1}{2}\frac{(\sigma+2f'\sigma')}{u}\,+\,\frac{1}{8}\frac{(3p- 
\rho)}{u}\,+\,\frac{3}{16}\frac{(\rho+p)}{f}\,,{\label{Delta
n}}
\end{equation}
\begin{equation}
\phi'=\,f'\frac{(3p-\rho)}{u}\,+\,\frac{(3f\sigma'-4\sigma f')}{u}\,, 
{\label{Delta phi}}
\end{equation}
where we have introduced
\begin{equation}
u\equiv 2\left[\,3f\,+\,8(f')^2\,\right]\,.{\label{U}}
\end{equation}
In the case that the brane tension does not depend on $\phi(t)$,  
these equations simplify to
\begin{equation}\frac{a'}{a}\,=\,\frac{1}{8}\frac{(3p-\rho-4\sigma)} 
{u}\,-\frac{1}{16}\frac{(\rho+p)}{f}\,,
\end{equation}
\begin{equation}n'\,=\,\frac{1}{8}\frac{(3p-\rho-4\sigma)}{u}\,+\, 
\frac{3}{16}\frac{(\rho+p)}{f}\,,
\end{equation}
\begin{equation}\phi'\,=\,f'\frac{(3p-\rho-4\sigma)}{u}\,.\end{equation}
\newpage

\section{Evolution Equations on the Brane}

 From equation ({\ref{05}}) we may obtain on the brane a {\textit 
{generalized continuity equation}}
\begin{equation}
\dot{\rho}+3\frac{\dot{a}}{a}(\rho+p)+2P_5=0 {\label{conti}}
\end{equation}
that expresses energy conservation. Notice that this equation is $f$-independent. From equation ({\ref{55}})
 we may obtain a {\textit{second order}} or {\textit{generalized Friedmann
equation}}
$$
3f\left[\,\frac{\ddot{a}}{a}+\left(\frac{\dot{a}}{a}\right)^2+\frac{k} 
{a^2}\right]\,+\,\ddot{f}\,+
\,3\frac{\dot{a}}{a}\dot{f}+\frac{1}{4}(\dot{\phi})^2-\frac{1}{2}\left 
(V+\Lambda\right)\,=\,
\frac{1}{64u}\left(3p-\rho-4\sigma\right)^2$$
\begin{equation}
\,-\frac{6}{(16)^2}\frac{(\rho+p)^2}{f}\,-\frac{1}{2}\overline{P}_B\,
-\frac{f'\sigma'}{4u}(3p-\rho-4\sigma)\,-\frac{3f}{8u}(\sigma')^2  
{\label{secondfried}}
\end{equation}
that expresses the time evolution of the scale factor $a(t)$. We have  
taken $k=0$. Finally, the scalar equation ({\ref{scalareq}}) gives on  
the brane
$$\ddot{\phi}+3\left(\frac{\dot{a}}{a}\right)\dot{\phi}+\frac{dV}{d 
\phi}-6f'\left[\frac{\ddot{a}}{a}+
\left(\frac{\dot{a}}{a}\right)^2\right]-\phi'\left(n'+3\frac{a'}{a} 
\right)$$
\begin{equation}\,
+\,6f'\left(\frac{a'}{a}\right)\left(\frac{a'}{a}+n'\right)
\,=\,\hat{\phi}''-2f'\left(3\frac{\hat{a}''}{a}+\hat{n}''\right)\,,
\end{equation}
where $\hat{\phi}''(t),\,\hat{a}''(t),\,\hat{n}''(t)$ stand for the
{\textit{non-distributional}} parts of these derivatives. The  
quantities $\hat{a}''$ and $\hat{n}''$ appear in
equations ({\ref{00}}), ({\ref{ij}}) considered on the brane and can  
be expressed in terms of $\hat{\phi}''$ and standard quantities, i.e.  
$a(t),\,\phi(t)$,
their time-derivatives and the various matter densities. 
Doing that\footnote{For simplicity we have taken $\sigma'=0$.}, we
obtain the set of three independent equations, namely ({\ref{conti}})  
and
$$\frac{\ddot{a}}{a}+\left(\frac{\dot{a}}{a}\right)^2\,\approx\,\frac 
{\sigma^2}{12
fu}\,+\,\frac{\sigma}{24 fu}\left(\rho-3p\right)\,-\frac{1}{6f} 
\overline{P}_B\,$$
\begin{equation}\,+\,\frac{1}{6f}(V+\Lambda)-\frac{\dot{\phi}^2}{12f} 
{{\left(1+4f''\right)}}
-\frac{f'}{3f}\left(\ddot{\phi}\,+\,3\frac{\dot{a}}{a}\dot{\phi} 
\right)\,,{\label{eqa}}
\end{equation}
$$\,$$
$$\ddot{\phi}+3\frac{\dot{a}}{a}\dot{\phi}-\hat{\phi}''+\frac{u'}{2u} 
\dot{\phi}^2\,\approx\,-\frac{6f}{u}V'(\phi)-\frac{2f'}{u}\left(3P_B+ 
\overline{P}_B-\rho_B\right)$$
\begin{equation}
\,+\,10\frac{f'}{u}(V+\Lambda)\,+\,{{4\frac{f'(u+f'u')\sigma}{u^3} 
\left(\rho-3p\right)}} \,+
\,{{\frac{8f'(u+f'u')\sigma^2}{u^3}}}\,.{\label{eqaphi}}
\end{equation}
We have enforced the {\textit{low energy density approximation}},  
where we neglect $\rho^2$ terms. Note that the scalar
evolution equation contains the unknown quantity $\hat{\phi}''$.
Although, up to now, we have worked in the framework of a general  
coupling function $f(\phi)$, we know that, since
$\phi=\phi(t)$, this cannot be very far from a constant at late times  
(Einstein gravity). Thus, we may consider a quadratic
coupling function
\begin{equation}
f(\phi)\,=\,2M^3\left(\,1-\frac{\xi}{2}\phi^2\,\right)\,,{\label 
{ffunction}}
\end{equation}
where $\xi$ is a suitable coupling parameter\cite{BD}\cite{KT}\cite 
{FP}. Note that ({\ref{ffunction}}) can be a good
approximation to a general coupling function $f(\phi)\approx \,f(0)\,+ 
\,f''(0)\phi^2/2\,+\,\cdots$ for small
$\phi<<(2f(0)/f''(0))^{1/2}$.

In order to get the scale factor evolution equation ({\ref{eqa}})  
into the familiar first order Friedmann equation form, we
may introduce an auxiliary variable $\chi(t)$, a so-called {\textit 
{dark energy field}}.
In the {\textit{dark energy field formulation}},\cite{KKTTZ},\cite{BT}
equation ({\ref{eqa}}) is replaced by a pair of two first order  
equations. These equations are
\begin{equation}
\left(\frac{\dot{a}}{a}\right)^2\,=\,2\gamma\rho\,+\,\chi\,+\,\lambda 
\,,{\label{Fried}}
\end{equation}
$$\,$$
$$\dot{\chi}\,+\,4\frac{\dot{a}}{a}\left\{\,\chi\,+\,\frac{1}{12f} 
\left[\,\,\overline{P}_B\,+\,
\frac{1}{2}(1-4\xi)\dot{\phi}^2\,-V\,-2\xi\phi\left(\ddot{\phi}\,+\,
3\frac{\dot{a}}{a}\dot{\phi}\,\right)\,\right]\,\,\right\}\,$$
\begin{equation}\,=\,4\gamma P_5\,-2\dot{\gamma}\rho\,-\dot{\lambda} 
\,.{\label{XI}}\end{equation}
Note that we have restricted ourselves to the ({\ref{ffunction}})  
form for the coupling function $f$, although we have not
substituted this expression everywhere. The functions $\beta,\,\gamma 
$ and $\lambda$ appearing in ({\ref{Fried}}) and
({\ref{XI}}) are defined as
\begin{equation}
\beta\equiv\frac{1}{24fu},\,\,\gamma\equiv \sigma \beta,\,\,\lambda 
\equiv
\frac{1}{12f}\left(\,\Lambda\,+\,\frac{\sigma^2}{2u}\,\right)\,.{\label{abc}}
\end{equation}
 The  
original second order differential equation
({\ref{eqa}}) can be recovered by differentiating ({\ref{Fried}}),  
inserting ({\ref{XI}}) and using the continuity equation
({\ref{conti}}) which is always assumed to hold. An equation  
expressing $\ddot{a}/a$ in terms of the dark energy field can
be written down, namely
\begin{equation}
\frac{\ddot{a}}{a}\,=\,-\gamma(\rho+3p)\,+\,\frac{1}{6f}\left[\,V\,- 
\overline{P}_B\,-\frac{1}{2}(1-4\xi)\dot{\phi}^2\,+\,2\xi\phi\left 
(\ddot{\phi}+
3\frac{\dot{a}}{a}\dot{\phi}\right)\,\right]\,-\chi\,+\,\lambda
\,.{\label{secondderiv}}
\end{equation}
The function $\lambda$ is the effective cosmological constant on the  
brane and, in the minimal case ($f=2M^3$),
the condition
 $$\lambda=\Lambda+\frac{\sigma^2}{24M^3}\,=0$$
  corresponds to the  
familiar Randall-Sundrum fine-tuning of {\textit{no
cosmological constant on the brane}} in the absence of matter.

\section{Late Time Behaviour in the Absence of Matter}
Let us now consider the case of absence of any matter. In this case,  
our set of equations becomes
\begin{equation}
\left(\frac{\dot{a}}{a}\right)^2\,=\,\chi\,+\,\lambda\,,{\label{darkfried}}
\end{equation}
$$\,$$
\begin{equation}\dot{\chi}\,+\,4\frac{\dot{a}}{a}\left\{\,\chi\,+\, 
\frac{1}{12f}\left[\,
\frac{1}{2}(1-4\xi)\dot{\phi}^2\,-V\,-2\xi\phi\left(\ddot{\phi}\,+\,
3\frac{\dot{a}}{a}\dot{\phi}\,\right)\,\right]\,\,\right\}\,=\,-\dot 
{\lambda}\,,{\label{dark}}\end{equation}
$$\,$$
\begin{equation}\ddot{\phi}+3\frac{\dot{a}}{a}\dot{\phi}-\hat{\phi}''+ 
\frac{u'}{2u}\dot{\phi}^2\,= \,-\frac{6f}{u}V'(\phi)\,+\,10\frac{f'} 
{u}(V+\Lambda)\,+
\,{{\frac{8f'(u+f'u')\sigma^2}{u^3}}}\,.{\label{field1}}
\end{equation}
Notice that in the minimal case ($f'=0$) the scalar equation can be  
satisfied with a time-independent solution
$\phi_0$ of $V'(\phi_0)=\hat{\phi}''$, provided $\hat{\phi}''$ is
time-independent as well. In this case, the dark energy equation  
({\ref{dark}}) is just ($V_0\equiv V(\phi_0)$)
\begin{equation}\dot{\chi}\,+\,4\frac{\dot{a}}{a}\left(\,\chi\,-\frac 
{V_0}{24M^3}\,\right)\,=\,0\,,
\end{equation}
with a solution
\begin{equation}
\chi\,=\,\frac{{\cal{C}}}{a^4}\,+\,\frac{V_0}{24M^3}\,.{\label{solution}}
\end{equation}
Finally, the effective Friedmann equation is
\begin{equation}
\left(\frac{\dot{a}}{a}\right)^2\,=\,\chi\,+\,\lambda\,=\,\frac{{\cal 
{C}}}{a^4}\,+\frac{V_0}{24M^3}\,+
\,\frac{1}{24M^3}\left(\,\Lambda\,+\,\frac{\sigma^2}{24M^3}\,\right) 
\,.{\label{solution-2}}\end{equation}
We can always choose $V_0=0$. Then, the condition for a vanishing  
cosmological constant on the brane takes the familiar Randall-Sundrum fine-tuning form $\Lambda+ 
\sigma^2/24M^3=0$. In the case of vanishing cosmological
constant the resulting
{\textit{dark radiation}}-dominated expansion is decelerating like $a(t)\propto t^ 
{1/2}$. Actually a constant solution is possible in the presence of the
non-minimal coupling as well and the above behaviour is not modified. Simply,
the condition on the constant $\hat{\phi}''$ is replaced by the more complicated
condition
$\hat{\phi}''= \,6fV'(\phi)/u\,-10f'(V+\Lambda)/u\,-8f'(u+f'u')\sigma^2/u^3$.
The dark energy field is now $\chi\,=\,{\cal{C}}/a^4\,+\,V_0/12f$, while the
Friedmann equation retains the same form with $\lambda$ defined in
({\ref{abc}}). Depending on the fine-tuning condition that we impose, we either
have a cosmological constant dominated exponential expansion or a dark energy
dominated $a(t)\propto t^{1/2}$ expansion as in the minimal case. Since the effective cosmological constant $\lambda$ depends on both $f$ and $u$, its value at late times will contain the coupling parameter $\xi$. A residual cosmological constant will thus be present, even if we impose the Randall-Sundrum fine-tuning, due to the non-minimal coupling, as can be seen from the expanded Friedmann equation
\begin{equation}
H^2  = \frac{\cal{C}}{{a^4 }} + \frac{{V_0 }}{{24M^3 }} + \xi \frac{{\phi ^2 }}{{48M^3 }}\left( {V_0  + \frac{{\sigma ^2 }}{{24M^3 }}\left( {1 - \frac{\xi }{{\xi _c }}} \right)} \right)
\end{equation}
where we only kept terms up to $\phi^2$ and imposed the fine-tuning. Apparently, the presence of the $V_0$ term is not necessary to get non-trivial results, except from the case of conformal coupling, $\xi=\xi_c=3/32M^3$.
 
Instead of viewing the scalar equation ({\ref{field1}}) as an equation that
supplies us with a solution for $\phi(t)$ for a given function $\hat{\phi}''$,
we may alternatively view it as an equation that determines the unknown function
$\hat{\phi}''$ in terms of a chosen configuration $\phi(t)$. Thus, we may choose
an exponentially decaying field
configuration 
\begin{equation}
\phi(t)\,=\,\phi_0\,e^{-\kappa t}{\label{exponential}}
\end{equation}
and assume that it is a solution of ({\ref{field1}}) for a suitable function $\hat{\phi}''$. Substituting it into
({\ref{dark}}), we obtain
\begin{equation}\dot{\chi}\,+\,4\frac{\dot{a}}{a}\left\{\,\chi\,+\, 
\frac{1}{12f}\left[\,
\frac{\kappa^2}{2}(1-4\xi)\phi^2\,-V\,-2\xi\phi^2\left(\kappa^2\,-
3\kappa\frac{\dot{a}}{a}\,\right)\,\right]\,\,\right\}\,=\,-\dot 
{\lambda}\,,{\label{expodark}}
\end{equation}
or, using ({\ref{darkfried}}),
\begin{equation}
\dot{\chi}\,+\,\dot{\lambda}\,+\,4\frac{\dot{a}}{a}\left\{\,\chi\,+\,\frac{\kappa^2(1-8\xi)}{24f}\phi^2\,-\frac{V}{12f}\,\right\}\,
+\,2\frac{\kappa\xi}{f}(\chi+\lambda)\phi^2\,=\,0\,.
\end{equation}
At late times, we may expand the scalar potential in powers of $\phi=\phi_0e^{-\kappa t}$ as
$$V(\phi)\,\approx\,V(0)\,+\,\frac{1}{2}V''(0)\phi^2+\dots\,,$$
assuming that it is a function of $\phi^2$. Thus, keeping only first order terms in $\phi^2$, we obtain
$$\dot{\overline{\chi}}\,+\,
4\frac{\dot{a}}{a}\left\{\,\overline{\chi}\,+
\,\frac{\phi^2}{48M^3}\left[-\xi\Lambda\,-\,\frac{\xi(1-\xi/2\xi_c)\sigma^2}{
12M^3 }\,+\,\kappa^2(1-8\xi)-V''(0)-\xi
V(0)\,\right]\right.$$
\begin{equation}\left.\,-\lambda(0)\,-\frac{V(0)}{24M^3}\,\right\}\,+
\,\frac{\kappa\xi}{M^3}\phi^2\overline{\chi}\,\approx\,0\,,
\end{equation}
where $\overline{\chi}=\chi+\lambda$ and
$\lambda(0)=(\Lambda+\sigma^2/24M^3)/24M^3$. To zeroth order, we have
$$\dot{\overline{\chi}}\,+\,
4\frac{\dot{a}}{a}\left(\,\overline{\chi}\,-\lambda(0)\,-\frac{V(0)}{24M^3}\,\right)\,\approx\,0\,,$$
with a solution identical to ({\ref{solution}}). The resulting Friedmann equation is just ({\ref{solution-2}}). Thus, at
late times, 
depending on the fine-tuning condition imposed, we either have exponential expansion driven by a cosmological constant
or a dark radiation driven decelerated expansion. The non-minimal coupling does not play any role to this order. Corrections which depend on $\xi$ occur once we include terms of order $\phi^2$.
In any case, the existence of the exponential solution ({\ref{exponential}}) rests on the scalar equation and
ultimately on $\hat{\phi}''$. Expanding it in powers of $\phi$ and substituting, we get
$$-\hat{\phi}''(0)\,+\,\left[\, \kappa^2-3\kappa
H(0)\,-\{\hat{\phi}''(0)\}_1\,+\,V''(0)\,+\,\frac{5\xi}{3}
\left(V(0)+\Lambda\right)\,+\frac{\xi} 
{9M^3}\sigma^2\,\right]\phi\,\,\,\,+\,$$
\begin{equation}
\left[\,-3\kappa H'(0)\,-\frac{1}{2}\{\hat{\phi}''(0)\}_2\,\right]\phi^2\,+\,\dots\,\approx\,0\,,{\label{vathmwti}}
\end{equation}
where we have introduced the Hubble parameter $H=\dot{a}/a$. We have introduced the expansions 
$$H \approx H(0)+H'(0)\phi\,+\,\dots$$
and
$$\hat{\phi}''\,\approx
\hat{\phi}''(0)\,+\,\{\hat{\phi}''(0)\}_1\phi\,+\,\frac{1}{2}\{\hat{\phi}''(0)\}_2\phi^2\,+\,\dots\,.$$
Equation ({\ref{vathmwti}}) is satisfied for $\hat{\phi}''(0)=0$,
$$\{\hat{\phi}''(0)\}_1\,=\,\kappa^2-3\kappa
H(0)\,+\,V''(0)\,+\,\frac{5\xi}{3}\left(V(0)+\Lambda\right)\,+\frac{\xi} 
{9M^3}\sigma^2$$
and
$$\{\hat{\phi}''(0)\}_2\,=\,-6\kappa H'(0)\,.$$
If in addition we restrict the forms of ${\hat \phi ''}$ and $V$, we would be forced to fine-tune the parameters $\kappa$, $H(0)$ and $H'(0)$. For ${\hat \phi ''}=0$ and $V=0$, equation (\ref{vathmwti}) reduces to
\begin{equation}
H'\left( 0 \right) = 0,\,\,\, H\left( 0 \right) = \frac{1}{{3\kappa}}\left( {\kappa^2  + \frac{{5\xi }}{3}\Lambda  + \frac{\xi }{{9M^3 }}\sigma ^2 } \right)
\end{equation}
which provides an approximate solution of constant H, which depends on $\xi$.
\section{A Class of Approximate Solutions}

Let us consider now our set of cosmological evolution equations in the  
presence of brane matter but in the absence of bulk matter, apart from the scalar field.
For the late-time behaviour of the scalar field we shall adopt here a decreasing power law
ansatz with an as yet unspecified power $\alpha$, namely
\begin{equation}
\phi(t)\,\approx\,\frac{C_1}{t^{\alpha}}\,\,\,.{\label{fiansatz}}
\end{equation}
We may use the original second order scale factor differential
equation instead of the dark energy field formulation since they are both equivalent.
 For the brane matter we shall assume the equation of
state $p=w\rho$. With this equation of state, the continuity equation ({\ref{conti}}) gives
\begin{equation}
\dot{\rho}+3\frac{\dot{a}}{a}(1+w)\rho\,=\,0\,\,\,\,\,\Longrightarrow 
\,\,\rho\,\propto\,a^{-3(1+w)}\,.{\label{Energy}}
\end{equation}
For the above ansatz our pair of evolution equations gives
$$\frac{\ddot{a}}{a}+\left(\frac{\dot{a}}{a}\right)^2\,\approx\,\frac 
{\sigma^2}{12
fu}\,+\,\frac{\sigma}{24 fu}\rho(1-3w)\,$$
\begin{equation}\,+\,\frac{1}{6f}(V+\Lambda)-\frac{\dot{\phi}^2}{12f} 
{{\left(1+4f''\right)}}
-\frac{f'}{3f}\left(\ddot{\phi}\,+\,3\frac{\dot{a}}{a}\dot{\phi} 
\right)\,,{\label{evol1}}
\end{equation}
$$\,$$
$$\ddot{\phi}+3\frac{\dot{a}}{a}\dot{\phi}-\hat{\phi}''+\frac{u'}{2u} 
\dot{\phi}^2\,\approx\,-\frac{6f}{u}V'(\phi)$$
\begin{equation}
\,+\,10\frac{f'}{u}(V+\Lambda)\,+\,{{4\frac{f'(u+f'u')\sigma}{u^3}\rho 
(1-3w)}} \,+
\,{{\frac{8f'(u+f'u')\sigma^2}{u^3}}}\,.{\label{evol2}}
\end{equation}
For the chosen quadratic form of the coupling function ({\ref{ffunction}}), $u$ takes on the simple form 
$u\,=\,12M^3\left[\,1\,-\xi\left(1-\xi/\xi_c\right) \phi^2/2\,\right]\,$, where $\xi_c\equiv 3/32M^3$ is the conformal value
of the non-minimal coupling parameter
\footnote{The effective cosmological {\textit{``constant"}}
 on the brane is $\lambda = \left(\Lambda  + \sigma ^2 /2u\,\right)/12f$. For
the conformal value of the 
 non-minimal coupling parameter $\xi=\xi_c$,
  we have $u=12M^3$ valid for all $\phi$ and the usual Randall-Sundrum 
  fine-tuning gives $\lambda=0$ as in the minimal case.}. 
  The coupling function derivatives that appear in the above equations are $f'=-(2M^3)\xi\phi$ and 
$u'=-12M^3(1-\xi/\xi_c )\xi\phi$. We shall also assume a simple quadratic scalar potential and without loss  
of generality we shall take it to be vanishing at the origin, i.e. $V(\phi)\,=\,\mu^2\phi^2/2\,$. 
 
Next, let us introduce an expanding scale factor ansatz
\begin{equation}
a(t)\,\approx\,C_3\,t^{\nu}\,,{\label{aansatz}}
\end{equation}
valid at late times, in terms of a power $\nu$ to be determined. Inserting the scale factor ansatz ({\ref{aansatz}}), 
the scalar field ansatz ({\ref{fiansatz}}) and substituting the energy density ({\ref{Energy}})
 as 
 \begin{equation}
 \rho \,=\,C_4\,t^{-3\nu(1+w)}\,
 \end{equation}
  into the  
Friedmann equation, we obtain
$$\frac{\nu(2\nu-1)}{t^2}\,\approx\,\frac{1}{48M^3}\left\{\,\frac 
{\sigma^2}{6M^3}+4\Lambda\,+\,\left(\frac{\sigma
(1-3w)C_4}{12M^3}\right)\frac{1}{t^{3\nu(1+w)}}\,\right.$$
$$+\,\frac{C_1^2}{t^{2\alpha}}\left[\,2\mu^2\,+2\xi\left(\Lambda\,+ 
\frac{4(2\xi_c-\xi)\sigma^2}{9}\right)\,
+ \frac{4\xi C_4(1-3w)(2\xi_c-\xi)\sigma}{9}\frac{1}{t^{3\nu(1+w)}}
\,\right.$$
\begin{equation}\left.\left.-\frac{2\alpha}{t^{2}}\left[\, \alpha(1-16M^3\xi)\,+\, 
8M^3\xi(3 \nu-1) \,\right]\,\,\right]\,\,\right\}\,.\end{equation}
 We assume that $\alpha>1$. Keeping only terms up to second order, we obtain the conditions
\begin{equation}\nu\,=\,\frac{2}{3(1+w)}\,,{\label{NU}}\end{equation}
\begin{equation}\Lambda\,+\,\frac{\sigma^2}{24M^3}\,=\,0\,,{\label{FT}}\end 
{equation}
\begin{equation}\nu(2\nu-1)\,=\,\frac{\sigma(1-3w)}{144(2M^3)^2}C_4\,. 
{\label{C4}}\end{equation}
Note that ({\ref{FT}}) is the standard fine-tuning for a vanishing cosmological constant on the brane. 

In a similar way the scalar equation gives
$$\alpha\left[\alpha +1-3\nu\right]\frac{C_1}{t^{\alpha +2}}\,-\hat 
{\phi}''\,+\,O(t^{-(3\alpha +2)})\,\approx\,
-\frac{\mu^2 C_1}{t^{\alpha}}\left[\,1\,-\frac{\xi^2}{2\xi_c}\frac 
{C_1^2}{t^{2\alpha}}\,\right]$$
$$-\frac{5\xi}{3}\,\frac{C_1}{t^{\alpha}}\,\left[\,\Lambda\,+\,\left 
(\,\frac{\mu^2}{2}\,+
\,\frac{\xi}{2}\left(1-\frac{\xi}{\xi_c}\right)\Lambda\,\right)\frac 
{C_1^2}{t^{2\alpha}}\,\right]\,$$
\begin{equation}-\frac{\xi \sigma}{18M^3}\left(2\sigma\,+\,\frac{C_4}{t^2}(1-3w) 
\right)\frac{C_1}{t^{\alpha}}\left[\,
1\,+\,\xi\left(1-\frac{\xi}{\xi_c}\right)\left(1\,+\,2M^3\xi\,\right) 
\frac{C_1^2}{t^{2\alpha}}\,\right]\,.{\label{EXISWSI-PHI}}\end{equation}
We have used equation ({\ref{NU}}). At this point we have to assume  
an ansatz for the unknown function $\hat{\phi}''$.

{\textit{\textbf{Ansatz 1 : $\bf{\hat{\phi}''\propto \ddot{\phi}}$}}}

This amounts to a late time behaviour
\begin{equation}\hat{\phi}''\,=\,\frac{C_2}{t^{\alpha +2}}\,.
\end{equation}
Substituting this into  ({\ref{EXISWSI-PHI}}), we obtain the two relations
\begin{equation}
C_2\,-\frac{(1-3w)\xi \sigma}{18M^3}C_1C_4\,=\,C_1\alpha\left(\, 
\alpha +1-3\nu\right)\,,\,\,\,\,\mu^2\,=\,-\frac{\xi \sigma^2}{24M^3}\,.
{\label{MU}}\end{equation}
We have considered terms only up to second order and ignored  
terms of $O(t^{-3\alpha})$.

 From equations ({\ref{NU}}), ({\ref{C4}}), ({\ref{MU}})
  we may fix $\Lambda,\,\mu^2$ and $C_4$ and obtain
\begin{equation}
\alpha=\frac{{3\nu  - 1}}{2} \pm \sqrt {\frac{{\left( {3\nu  - 1}  
\right)^2 }}{2} + \frac{{C_2 }}{{C_1 }} - 3\frac{\xi }{{\xi _c }}\nu  
\left( {2\nu  - 1} \right)} \,.
\end{equation}
Note that $\nu$ is given by ({\ref{NU}}). For $\nu<1$, or  
equivalently, $w>-1/3$, the exponent $\alpha$ is real for
any value of $\xi$. In contrast, for $\nu>1$, or equivalently,  
$-1<w<-1/3$, the exponent $\alpha$ is real and positive
for
$$\xi\,\leq\, \xi_c\frac{\left[(3\nu-1)^2+4\frac{C_2}{C_1}\right]}{12\nu(\nu-1)}\,.$$
This inequality restricts the values that $\xi$ is allowed to take in
order to have real values for the exponent $\alpha$ and a value of $\nu$ corresponding to accelerated expansion. The  
presence of
the free parameter $\frac{C_2}{C_1}$ allows a certain degree of flexibility in  
the choice
of $\xi$ for small values of $\nu$.

{\textit{\textbf{Ansatz 2 : $\bf{\hat{\phi}''\propto \phi}$}}}

This amounts to a late time behaviour
\begin{equation}
\hat{\phi}''\,=\,\frac{C_2}{t^{\alpha}}\,,
\end{equation}
Substituting this into the scalar equation ({\ref{EXISWSI-PHI}}), we obtain
$$\alpha\left[\alpha +1-3\nu\right]\frac{C_1}{t^{\alpha +2}}\,-\frac 
{C_2}{t^{\alpha}}\,+\,O(t^{-(3\alpha +2)})\,\approx\,
-\frac{\mu^2 C_1}{t^{\alpha}}\left[\,1\,-\frac{\xi^2}{2\xi_c}\frac 
{C_1^2}{t^{2\alpha}}\,\right]$$
$$-\frac{5\xi}{3}\,\frac{C_1}{t^{\alpha}}\,\left[\,\Lambda\,+\,\left 
(\frac{\mu^2}{2}\,+
\,\frac{\xi}{2}\left(1-\frac{\xi}{\xi_c}\right)\Lambda\,\right)\frac 
{C_1^2}{t^{2\alpha}}\,\right]\,$$
\begin{equation}-\frac{\xi \sigma}{18M^3}\left(2\sigma\,+\,\frac{C_4}{t^2}(1-3w) 
\right)\frac{C_1}{t^{\alpha}}\left[\,
1\,+\,\xi\left(1-\frac{\xi}{\xi_c}\right)\left(1\,+\,2M^3\xi\,\right) 
\frac{C_1^2}{t^{2\alpha}}\,\right]\,.\end{equation}
Neglecting terms of $O(t^{-3\alpha})$, we obtain the  
relations
\begin{equation}
\frac{C_2}{C_1}=\mu^2+\frac{5\xi}{3}\Lambda+\frac{\xi\sigma^2}{9M^3} 
\,,{\label{MU1}}
\end{equation}
\begin{equation}
\alpha(\alpha+1-3\nu)=-\frac{(1-3w)\xi\sigma}{18M^3}C_4\,.
\end{equation}
 From these relations we obtain an almost identical expression for the  
exponent $\alpha$ as in the case of the previous ansatz. The only difference is that the term
$C_2/C_1$ is missing, namely
\begin{equation}
\alpha\,=\,\frac{3\nu-1}{2}\pm\sqrt{\frac{(3\nu-1)^2}{4}-3\frac{\xi} 
{\xi_c}\nu(2\nu-1)}
\,.
\end{equation}
Note that $\nu$ is given by ({\ref{NU}}). For $\nu<1$, or  
equivalently, $w>-1/3$, the exponent $\alpha$ is real for
any value of $\xi$. In contrast, for $\nu>1$, or equivalently,  
$-1<w<-1/3$, the exponent $\alpha$ is real and positive
for
\begin{equation}
\xi\,\leq\, \xi_c\frac{(3\nu-1)^2}{12\nu(\nu-1)}\,.{\label{XX}}
\end{equation}
In comparison to the previous ansatz, the range of allowed values for $\xi$ is
smaller here.
In this case the free parameter $\frac{C_2}{C_1}$ is present in the expression for $\mu^2$ obtained from ({\ref 
{MU1}}), namely
$$\mu^2\,=\,\frac{C_2}{C_1}\,-\,\frac{\xi\sigma^2}{24M^3}\,.$$

\section{Conclusions}

In the present article we studied the cosmological evolution on a brane embedded in a $5D$ bulk in 
the presence of a bulk scalar field non-minimally coupled to the Ricci curvature scalar through a term $f(\phi)R$. We derived the cosmological
evolution 
equations on the brane in the presence of arbitrary brane and bulk matter.
The scalar field evolution equation on the brane contains the non-distributional part of the second derivative of the
scalar field ($\hat{\phi}''$). The scalar evolution equation includes
an unknown non-distributional part of the second derivative of the scalar field with respect to the fifth coordinate, a
quantity that requires the knowledge of the dependence of the scalar field on the fifth coordinate. We have proceeded, 
considering this quantity as a function depending on the structure of the bulk and introduced various ans\"atze for it. 
Although we derived the evolution equations for a general coupling function, 
we focused on a quadratic form for it, namely $f(\phi)=2M^3(1-\xi\phi^2/2)$. We considered the 
\textit{``dark energy field"} formulation of the scale factor evolution equation in which the second order differential 
equation for the scale factor is replaced by an effective first order Friedmann equation and an equation for an auxiliary
variable. We first consider the case that no brane or bulk matter were present. 
We showed that late-time solutions exist, with 
a constant or exponentially decaying 
scalar field, in which the scale factor evolution is driven either by an effective cosmological constant or by dark
radiation. The value of the effective cosmological constant may in both cases depend on the coupling parameter. Although, no restriction is placed by these late-time solutions on the non-minimal coupling parameter, they
do depend on the behaviour of $\hat{\phi}''$. Next, we have considered the case of brane matter in addition to the scalar
field. We introduced an expanding power law ansatz for the scale factor ($a\sim t^{\nu}$)and 
assumed a decreasing power law type of time-dependence of the scalar field at late times ($\phi\sim t^{-\alpha}$). The
resulting constraints depend on the assumptions on the behaviour of $\hat{\phi}''$. Considering first the case
$\hat{\phi}''\propto \ddot{\phi}$, we see that we indeed have such a late-time solution only for brane matter with an
equation of state parameter $-1<w<-1/3$ and a non-minimal coupling parameter $\xi<\xi_c[(3\nu-1)^2+4C_2/C_1]/12\nu(\nu-1)$. 
The ratio $C_2/C_1$ is the $\hat{\phi}''/\phi$ coefficient ratio. We next considered the case $\hat{\phi}''\propto \phi$ and
found that, again, such a late-time solution only for brane matter with an
equation of state parameter $-1<w<-1/3$ and a non-minimal coupling parameter $\xi<\xi_c(3\nu-1)^2/12\nu(\nu-1)$. These
results correspond to a simple quadratic scalar potential choice but are
expected not to be modified by higher potential terms at late times.
Nevertheless, they could be modified by the presence of bulk matter interacting
with the brane. Depending on the ansatz for the bulk pressure and exchange
densities, the range for $w$ could be replaced by standard values.

\bigskip

{\textbf{ Acknowledgments.}}
This research was co-funded by the European Union
in the framework of the Program $\Pi Y\Theta A\Gamma O PA\Sigma-II$
of the {\textit{``Operational Program for Education and Initial  
Vocational Training"}} ($E\Pi EAEK$) of the 3rd Community Support  
Framework of the Hellenic Ministry of Education, funded by $25\% $ from
national sources and by $75\%$ from the European Social Fund (ESF).  
C. B. acknowledges also an {\textit{Onassis Foundation}} fellowship.

\end{document}